\documentclass[twoside,superscriptaddress,twocolumn,floatfix,a4paper,pra]{revtex4}

\usepackage{color}
\usepackage{graphicx}
\usepackage[utf8x]{inputenc}
\usepackage[T1]{fontenc}
\usepackage{siunitx}
\usepackage{amstext}
\usepackage{textgreek}
\usepackage{hyperref}
\usepackage{epstopdf}
\usepackage{amsfonts}
\usepackage{amsmath}
\usepackage{multirow}

\begin{document}

\title{Experimental characterization of photon-number noise in Rarity-Tapster type interferometers}

\author{Vojtěch Trávníček}
\email{vojtech.travnicek01@upol.cz}
\affiliation{RCPTM, Joint Laboratory of Optics of Palacký University and Institute of Physics of Czech Academy of Sciences, 17. listopadu 12, 771 46 Olomouc, Czech Republic}

\author{Karol Bartkiewicz} \email{bartkiewicz@jointlab.upol.cz}
\affiliation{Faculty of Physics, Adam Mickiewicz University,
PL-61-614 Pozna\'n, Poland}
\affiliation{RCPTM, Joint Laboratory of Optics of Palacký University and Institute of Physics of Czech Academy of Sciences, 17. listopadu 12, 771 46 Olomouc, Czech Republic}

\author{Antonín Černoch} \email{antonin.cernoch@upol.cz}
\affiliation{RCPTM, Joint Laboratory of Optics of Palacký University and Institute of Physics of Czech Academy of Sciences, 17. listopadu 12, 771 46 Olomouc, Czech Republic}
   
\author{Karel Lemr}
\email{k.lemr@upol.cz}
\affiliation{RCPTM, Joint Laboratory of Optics of Palacký University and Institute of Physics of Czech Academy of Sciences, 17. listopadu 12, 771 46 Olomouc, Czech Republic}   

\begin{abstract}
In this paper, we develop a simple model describing inherent photon-number noise in Rarity-Tapster type interferometers. This noise is caused by generating photon pairs in the process of spontaneous parametric down-conversion and adding a third photon by attenuating fundamental laser mode to single-photon level. We experimentally verify our model and present resulting signal to noise ratios as well as obtained three-photon generation rates as functions of various setup parameters. Subsequently we evaluate impact of this particular source of noise on quantum teleportation which is a key quantum information protocol using this interferometric configuration.
\end{abstract}

\date{\today}

\maketitle
\section{Introduction}
Quantum information processing (QIP) is a modern and perspective reaserch discipline of information science \cite{H11,H12,H13}. One of the platforms suitable for QIP are discrete photons manipulated using linear optics \cite{H21}. This platform is particularly promising for quantum communications, because of fast and relatively noiseless propagation of individual photons through open space or in fibers \cite{HP1,HP2}.

\begin{figure}[t]
		\begin{center}
		\includegraphics[scale=0.5]{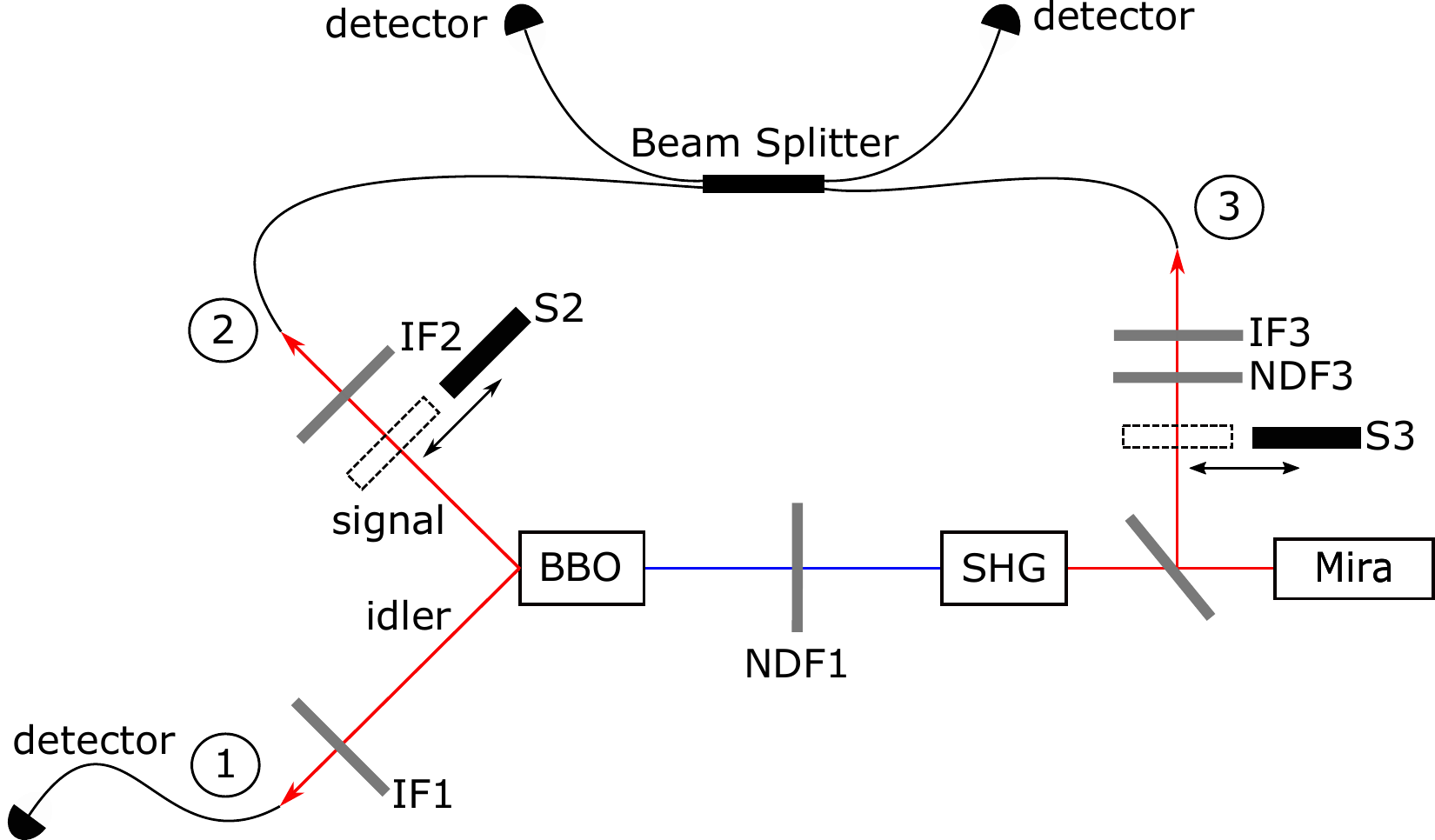}
		\caption{Setup of the experiment, 1 -- idler mode, 2 -- signal mode, 3 -- attenuated fundamental laser mode, IF(1-3) -- interference filters (3 nm in FWHM), NDF(1,3) -- neutral density filter, S(2,3) -- shutters, SHG -- second harmonics generation, Mira -- Ti-sapphire fs laser (central wavelength of $826$ nm, FWHM of $11$ nm), BBO -- a $\beta$-BaB${}_2$O${}_4$ crystal for SPDC.}
		\label{setup}	
		\end{center}
\end{figure}

Quantum teleportation \cite{H84,H85} is a key ingredient for many quantum information protocols such as entanglement swapping \cite{E1}, quantum relays \cite{HP3} or teleportation-based quantum computing \cite{E2}. On the platform of linear optics, quantum teleportation is usually achieved in the so-called Rarity-Tapster interferometer \cite{TR} (shown in Fig. \ref{setup}). In this interferometer, one photon from an entangled pair gets overlapped on a balanced beam splitter with an independent photon \cite{H21}. The output ports of the beam splitter are then subjected to suitable Bell-state projection.
Mulitphoton inteferometers have also a number of potential
applications that go beyond quantum teleportation (for a review see Ref.~\cite{PanRMP}). For example, they can be also used for engineering cluster states~\cite{Tashima}.

Single--photon sources used in experimental quantum information processing today are however imperfect and the number of photons generated per pulse is random, given by the state's photopulse statistics (e.g. Bose-Einstein, Poisson). While vacuum states can be filtered out by suitable post-selection, higher photon-number contributions can not always be recognized \cite{E3,E4}.

In 1988, Ou and Mandel predicted that visibility of two-photon bunching with classical beams is limited to 50\% due to their photon-number statistics \cite{OM88}. This research was further generalized to interaction between classical beam and ideal single-photon source \cite{TR}. Subsequently, researchers have managed to considerably increase visibility in Rarity-Tapster interferometers by optimizing spectral properties of interacting beams \cite{L09,J11,H13b}. Independently, several research groups have investigated two-photon bunching between two heralded single-photon sources \cite{B14,J15,W16}.

In this paper, we present a simple and practical model describing inherent photon-number noise in Rarity-Tapster type interferometers based on sources using spontaneous parametric down conversion (SPDC) and attenuated coherent state. These are currently predominant photon sources in experimental linear-optical QIP \cite{HP1,H58,HA11,HA7,E5,E6,E7}. We have experimentally tested validity of our model and established both theoretical and experimental relations between photon-number noise and various setup parameters. {Our goal was to investigate the effect of photon--number noise originating directly in photon sources.} To our best knowledge no article providing such analysis has yet been published. The influence of transmission noise on the fidelity and security of quantum teleportation of qubits was analyzed in Ref.~\cite{Sahin}.
Photon-number noise does not originate from experimental imperfections but is rather an intrinsic property of various photon sources (having their photon-number statistics). This fact even further stresses out the importance of this investigation.

The paper is organized as follows: In Sec. \ref{th model} we develop a theoretical model describing dependency of signal--to--noise ratio on the main parameters of the experimental setup. In Sec. \ref{experiment} we present experimnetal data verifing our model. In Sec. \ref{sec4} we investigate the impact of the photon--number noise on teleporation fidelity. We conclude in Sec. \ref{conclusions}.

\section{Theoretical model}
\label{th model}
%\emph{Theoretical model} --- 
Here, we assume that the pairs of photons are generated in the process of degenerate parametric down conversion.  
The generated optical fields are not strictly monochromatic, but for each wavelength from their spectrum the following reasoning holds. 
Let us denote $|\psi_{s}\rangle$ the state of signal and idler modes of the SPDC generated photons (Nos. 1 and 2) and $|\alpha\rangle$ the coherent state of the attenuated fundamental laser mode (No. 3).
We start with the Hamiltonian for SPDC process in the form of \cite{hamiltonian}
\begin{equation}
\hat{H}_{\mathrm{SPDC}} = \gamma \alpha_{p} \hat{a}_{1}^\dagger \hat{a}_{2}^\dagger + h.c. ,
\end{equation}
where $\gamma \ll 1$ is an interaction constant, $\alpha_{p}$ is a strong pumping amplitude of frequency doubled laser beam and $\hat{a}_{1}^\dagger$, $\hat{a}_{2}^\dagger$ are creation operators of the idler and signal photon modes respectively. The corresponding evolution operator is then of the form of
\begin{equation}
\hat{U} = \mathrm{exp}\left(\dfrac{i}{\hbar}\hat{H} t \right).
\end{equation}
%We can aproximate this evolution operator by first three terms of its Taylor expansion
%\begin{equation}
%\hat{U} \approx 1 + \dfrac{i \hat{H} t}{\hbar} + \Big(\dfrac{i t}{\hbar}\Big)^2 \dfrac{\hat{H}^2}{2}.
%\end{equation}
The state of the signal and idler modes is obtained by applying the $\hat{U}$ operator to the initial vacuum state
\begin{equation}
|\psi_{s}\rangle \propto|0 0\rangle + \dfrac{it}{\hbar}\gamma\alpha_{p} |1 1\rangle + \frac{\left(i t \gamma \alpha_{p}\right)^2}{2\hbar^2}|2 2\rangle +...
\end{equation}
We can express this state as
\begin{equation}
|\psi_{s}\rangle \propto |0 0\rangle + \kappa |1 1\rangle + \dfrac{\kappa^2}{2}|2 2\rangle,
\label{eq:psis}
\end{equation}
for $|\kappa|\ll 1$ and
\begin{equation}
\kappa = \dfrac{it}{\hbar}\gamma\alpha_{p}.
\end{equation}
The term $|0 0\rangle$ in Eq.~(\ref{eq:psis}), can be omitted because the first photon works as a herald which means that if it does not get detected the measurement will not succeed. This is under the assumption of negligible dark counts.
Furthermore, we have to take into account probability of coupling the photons from SPDC into optical fibers. Let us denote $t_{1}$ and $t_{2}$ the amplitude coupling efficiency of idler and signal modes respectively. 
The state of the first and second photon then reads
\begin{eqnarray}
\label{eq:psis_t}
|\psi_{s}\rangle & \propto & 2\kappa t_{1} t_{2}|1 1\rangle + 2\kappa t_{1} \sqrt{1-t_{2}^2}|1 0\rangle +\nonumber\\
& + & \kappa^2 t_{1} \sqrt{1-t_1^2} t_{2}^2 |1 2\rangle + \kappa^2 t_{1}^2 t_{2}^2 |2 2\rangle,
\end{eqnarray}
where again we have excluded the terms corresponding to the first mode being in a vacuum state.
Moreover, the last term in Eq. (\ref{eq:psis_t}) can be neglected with respect to the third term since in typical experimental setups $t_{1,2} \ll 1$.

Next, we can express the coherent state of attenuated fundamental laser mode of the same wavelength as the generated photon pairs in Fock basis and limit the expansion to first $N$ terms
\begin{equation}
|\alpha\rangle \approx \sum_{n=0}^N \frac{\alpha^n}{\sqrt{n!}}|n\rangle.
\label{eq:coh_fock}
\end{equation}
Any filtering or coupling efficiency do not change the nature of the attenuated laser mode which remains in a coherent state with amplitude $\alpha$ already including all possible losses. Thus we do not need to consider its coupling efficiency like in the SPDC modes. 

If the source were to be perfect, there should be precisely one photon in each of the three modes. Simultaneous detection of these photons corresponds to genuine coincidences denoted $CC_{g}$. In reality, SPDC-based sources yield also higer-photon-number contributions. On the beam splitter, these photons may split leading to three-photon detection even if there were no photons in the attenuated laser mode [see the third term in Eq. (\ref{eq:psis_t})]. These detections denoted $CC_{s}$ contribute to added noise. Similar source of noise are higer photon-number contributions from the fundamental laser mode that again can split on the beam splitter resulting in parasitic detections $CC_{f}$. Using Eqs. (\ref{eq:psis_t}) and (\ref{eq:coh_fock}) for $N=3$, we can identify the generation probabilities of the genuine coincidences as well as of the two parasitic contributions
\begin{equation}
CC_{g} \propto |\kappa|^2 |\alpha|^2 t_{1}^2 t_{2}^2,
\label{eq:ccg}
\end{equation}
\begin{equation}
\label{eq:ccs}
CC_{s} \propto t_{1}^2 t_{2}^4\dfrac{|\kappa|^4}{4},
\end{equation}
\begin{equation}
\label{eq:ccf}
CC_{f} \propto |\kappa|^2 t_{1}^2 \left(\dfrac{ |\alpha|^4 }{2}+\dfrac{|\alpha|^6}{6}\right).
\end{equation}
Note that in Eq. (\ref{eq:ccs}), we have assumed $1-t_1^2 \approx 1$ and in Eq. (\ref{eq:ccf}) $1-t_2^2 \approx 1$. These approximation are valid especially when one considers a linear--optical setup fed by the source which strongly diminishes the transmisivity due to technological losses (back--scattering, fiber coupling etc.)

The goal now is to maximize the signal-to-noise ratio defined as
\begin{equation}
\mathrm{SNR} \equiv \dfrac{CC_{g}}{CC_{s}+CC_{f}} = \dfrac{12 |\alpha|^2 t_{2}^2}{3 |\kappa|^2 t_{2}^4 + 6 |\alpha|^4+ 2 |\alpha|^6} .
\end{equation}
In a typical setup as depicted in Fig. \ref{setup}, there are two parameters that can easily be tuned: (i) amplitude of the attenuated fundamental laser mode $\alpha$ and (ii) SPDC pumping amplitude $\alpha_{p}$. In subsequent analysis, we investigate the dependency of SNR on these two parameters.

First we look at SNR as function of $\alpha$, which translates to the observed ratio $R$ between coincidence rates $CC_{f}$ and $CC_{s}$
\begin{equation}
\label{eq:R}
R \equiv \dfrac{CC_{f}}{CC_{s}} = \dfrac{2|\alpha|^4}{|\kappa|^2 t_{2}^4} + \dfrac{2|\alpha|^6}{3|\kappa|^2 t_{2}^4} \approx \dfrac{2|\alpha|^4}{|\kappa|^2 t_{2}^4}.
\end{equation}
We have omitted the second expansion term from $CC_{f}$ because for typical levels of attenuation to single-photon level $|\alpha| \ll 1$. The signal-to-noise ratio can now be approximated as function of the parameter $R$
\begin{equation}
\mathrm{SNR} \approx  \frac{2\sqrt{2R}}{|\kappa|(R+1)}.
\label{max}
\end{equation}
One can now find optimal value of $R$ by searching for maximum of this function. When $|\alpha| \ll 1$ holds, the optimal value of $R$ is $1$. For larger values of $|\alpha|$ the optimal $R$ shifts to slightly lower values {because the approximation in Eq. (\ref{eq:R}) does not longer apply.} In an experiment, one should thus seek to balance the false coincidence rates from SPDC and from attenuated fundamental mode.

In the subsequent analysis, we assume that $|\alpha| \ll 1$ holds and fix the parameter $R$ at its optimal value of 1. The Eq. (\ref{max}) then simplifies into the form
\begin{equation}
\label{model2}
\mathrm{SNR} = \frac{2\sqrt{2}}{|\kappa|},
\end{equation}
which can, with the help of Eqs. (\ref{eq:ccg}) and (\ref{eq:R}), be expressed in terms of the genuine coincidence rate $CC_g$
\begin{equation}
\label{eq:SNR_CCg}
\mathrm{SNR} \propto \sqrt[3]{\frac{16t_1^2t_2^4}{CC_g}}.
\end{equation}
One can now make two important conclusions towards the performance of the interferometer. Firstly, the SNR can only be increased by decreasing the value of $|\kappa|$ which means by lowering the SPDC pumping strength $|\alpha_P|$. Secondly, the obtained coincidence rate depends on the coupling efficiency of the signal and idler SPDC modes. Especially, it scales with the fourth power of the amplitude transmissvity of the signal mode (or second power of intensity transmissivity). For any given pumping strength, one can improve the overall coincidence rate by improving the coupling efficiencies. The SNR, however, can not be improved by this adjustment.

\section{Experimental implementation}
\label{experiment}
%\emph{Experimental testing} --- 
We have subjected our model and the resulting conclusions to an experimental test. Our experimental setup is depicted in Fig. \ref{setup}. The attenuated fundamental laser mode (mode No. 3) is obtained by splitting a small portion from the femtosecond pumping laser beam (Coherent Mira at \SI{826}{\nano\meter}). It then passes through a neutral density filter (NDF3) and 3nm-wide interference filter (IF3) before been coupled into single-mode fiber. 

The main laser beam enters second harmonics generation unit (SHG), where its wavelength becomes \SI{413}{\nano\meter}. The beam then passes through a neutral density filter (NDF1) and enters a Type I cut BBO crystal (\SI{0.64}{\milli\meter} thick) which due to SPDC generates idler and signal photons (Nos. 1 and 2) respectively. The photons in signal mode then pass through a 3nm-wide interference filter (IF2). The photons in idler mode pass through a 10nm-wide interference filter (IF1). The two SPDC modes are then coupled into single-mode fibers, idler mode is directly lead to a single-photon detector unlike the modes 2 and 3 that are mixed in a 50:50 fiber coupler before being detected. The avalanche photodiode detectors with suitable electronics record three-fold coincidence detections. Coincidence detection window was set to \SI{5}{\nano\second}, less than the laser repetition period of approximately \SI{12.5}{\nano\second}. We set the temporal displacement between photons 2 and 3, so they do not overlap in the fiber coupler. Thus we prevent the effect of two-photon interference.

In our experiment, we performed all the testing measurements in three steps: (i) with the shutters S2 and S3 open we detect all three-fold coincidences $CC_{a}$ which include $CC_{g}$ and parasitic contributions from signal and attenuated fundamental laser mode $CC_{s}$ and $CC_{f}$
\begin{equation}
CC_{a} = CC_{g} + CC_{f} + CC_{s}.
\label{cca}
\end{equation}
(ii) then we close shutter S3 and obtain three-fold coincedences only if there is more than one photon in signal mode, thus we measure parasitic coincidence rate $CC_{s}$. (iii) finally we close shutter S2, open S3 and therefore obtain three-fold coincedences only if there is more than one photon in attenuated fundamental laser mode -- parasitic coincidence rate $CC_{f}$. Note that $CC_{g}$ is obtained from Eq. (\ref{cca}) simply by subtracting $CC_{f}$ and $CC_{s}$ from $CC_{a}$. Each step took about \SI{100}{\second} and the entire three-step procedure was reapeted multiple times, thus we have avioded a bias caused by long-term laser power fluctuations.

\begin{figure}[!htb]
		\begin{center}
		\includegraphics[scale=0.3]{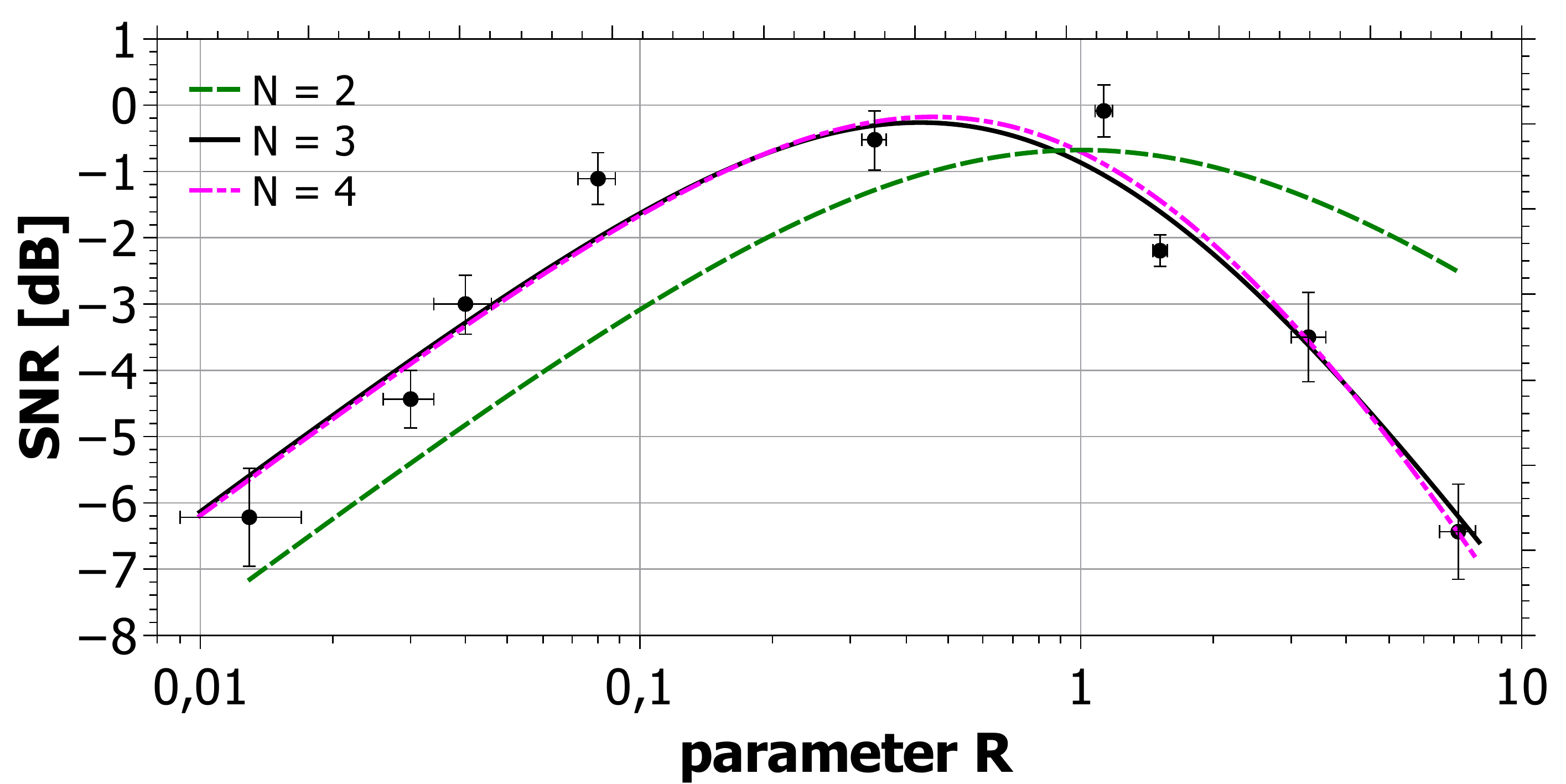}
		\caption{Dependence of SNR on parameter R. Points visualize experimentally observed results. {Lines correspond to various levels of expension in Eq. (\ref{eq:coh_fock}): to 2 (green dashed line), 3 (black solid line) and 4 (magneta dashed--dot line) terms.}}	
		\label{SNR1}	
		\end{center}
\end{figure}

First, we have experientially verified the dependence of SNR on $\alpha$, hence as a function of $R$ [see Eq. (\ref{max})]. The experiment consisted of measuring the coincidence rates for various values of R using the above-mentioned three steps. The parameter R was changed by modifying transmissivity of NDF3. Experimentally obtained values are summarized in Tab. \ref{tab1} and visualized in Fig. \ref{SNR1} together with the theoretical fit based on Eq. (\ref{max}). The dashed line shows a fit in which we limited the expansion in Eq. (\ref{eq:coh_fock}) to the first three terms, however it turns out that the model is not accurate enough for $R \to 10$ (see Fig. \ref{SNR1}). With growing contribution of parasitic coincidences from the attenuated fundamental laser mode $CC_{f}$, and thus also growing ratio $R$, higher terms in Eq. (\ref{eq:coh_fock}) can no longer be neglected {and the approximation in Eq. (\ref{eq:R}) does no longer hold}. The solid line which represents a model where we used the first four terms of the expansion, is accurate enough throughout the entire measured range of $R$. We went a step further and expended our model (represented in Fig. \ref{SNR1} by dash-dot line) to include the first five terms of the expansion. There is a slight but unsubstantial improvement to the previous case and thus we find the four-term expansion to be the optimum compromise between accuracy and complexity. {To simplify the following experiments, we have set the attenuated laser beam power so that the approximation in Eq. (\ref{eq:R}) holds. This means setting $\mathrm{R} \in [0.2;1]$ which also coincides with the SNR maximum.}

\begin{table}
\begin{ruledtabular}
\begin{tabular}{cc}
    \bfseries SNR [dB] & \bfseries parameter R\\[2mm]
    \hline
    -6.222 $\pm$ 0.740 & 0.013 $\pm$ 0.004  \\
    -4.440 $\pm$ 0.432 & 0.030 $\pm$ 0.004  \\
    -3.010 $\pm$ 0.440 & 0.040 $\pm$ 0.006  \\
    -1.105 $\pm$ 0.388 & 0.080 $\pm$ 0.008  \\
    -0.530 $\pm$ 0.442 & 0.340 $\pm$ 0.021  \\
    -0.086 $\pm$ 0.392 & 1.130 $\pm$ 0.052  \\
    -2.201 $\pm$ 0.241 & 1.510 $\pm$ 0.057  \\
    -3.502 $\pm$ 0.667 & 3.290 $\pm$ 0.290  \\
    -6.434 $\pm$ 0.727 & 7.180 $\pm$ 0.680  \\
\end{tabular}
\caption{Experimentaly observed data and their respective errors when investigating the dependence of SNR on the parameter $R$}
\label{tab1}
\end{ruledtabular}
\end{table}

\begin{figure}[!htb]
		\begin{center}
		\includegraphics[scale=0.3]{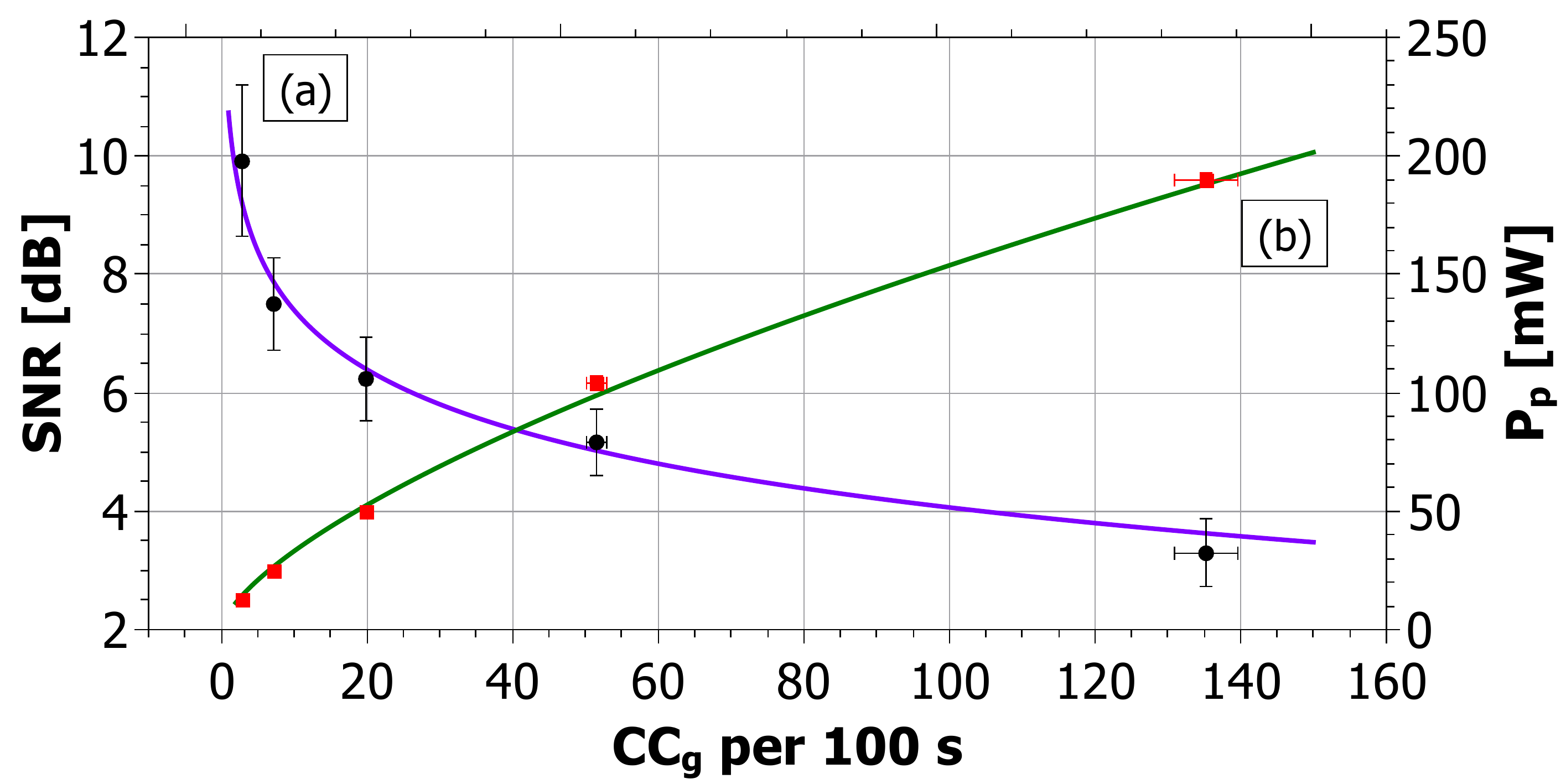}
		\caption{(a) Dependence of SNR on genuine coincidence rate $CC_{g}$. Points visualize experimentally observed results, the solid violet line depicts fitted experimental data with theoretical dependence based on Eq. (\ref{model2}). (b) Dependence of $CC_{g}$ on $P_{p}$.  The solid green line depicts fitted experimental data with theoretical dependence based on Eq. (\ref{eq:ccg}).}
		\label{SNR2}		
		\end{center}
\end{figure}

As the next test, we have measured the dependence of SNR on the pumping amplitude $\alpha_{p}$, which also translates into the dependence of SNR on the genuine coincidence rate $CC_{g}$ [see Eqs. (\ref{model2}) and (\ref{eq:SNR_CCg})]. We maintained the ratio $R$ close to its optimum discovered in previous test ($R \approx 0.35 \pm 0.04$) and were changing $\alpha_{p}$ by changing transmissivity of NDF1. So for every measured value of SNR, we have adjusted both the NDF1 (influencing $\alpha_{p}$) and NDF3 (to maintain constant $R$). The measurement procedure was also realised in the previously mentioned three acquisition steps. Experimentally obtained values are summarized in Tab. \ref{tab2} and visualized in Fig. \ref{SNR2} together with a theoretical fit based on Eq. (\ref{model2}). The Fiq. \ref{SNR2} proves that our four-term model matches well the experimental data. We have also investigated dependence of $CC_{g}$ on pumping power $P_{p}$ which is proportional to pumping amplitude $|\alpha_{p}|^2$.

\begin{table}
\begin{ruledtabular}
\begin{tabular}{ccc}
    \bfseries SNR [dB] & \bfseries CC${}_g$ per \SI{100}{\second}  & \bfseries  $P_{p}$ $\propto$ $|\alpha_{p}|^2$ [\si{\milli\watt}] \\[2mm]
    \hline
    9.91 $\pm$ 1.274 & 2.91 $\pm$ 0.111 & 13  $\pm$ 2  \\
    7.50 $\pm$ 0.787 & 7.23 $\pm$ 0.217 & 25  $\pm$ 2  \\
    6.23 $\pm$ 0.714 & 19.88 $\pm$ 0.613 & 50  $\pm$ 2  \\
    5.17 $\pm$ 0.559 & 51.59 $\pm$ 1.384 & 104  $\pm$ 3  \\
    3.33 $\pm$ 0.577 & 135.28 $\pm$ 4.392 & 190  $\pm$ 3  \\
\end{tabular}
\caption{Experimentally observed data and their respective errors when investigating the dependence of SNR on the $CC_{g}$ and $CC_{g}$ on the $\alpha_{p}$.}
\label{tab2}
\end{ruledtabular}
\end{table}

\begin{figure}[!htb]
		\begin{center}
		\includegraphics[scale=0.3]{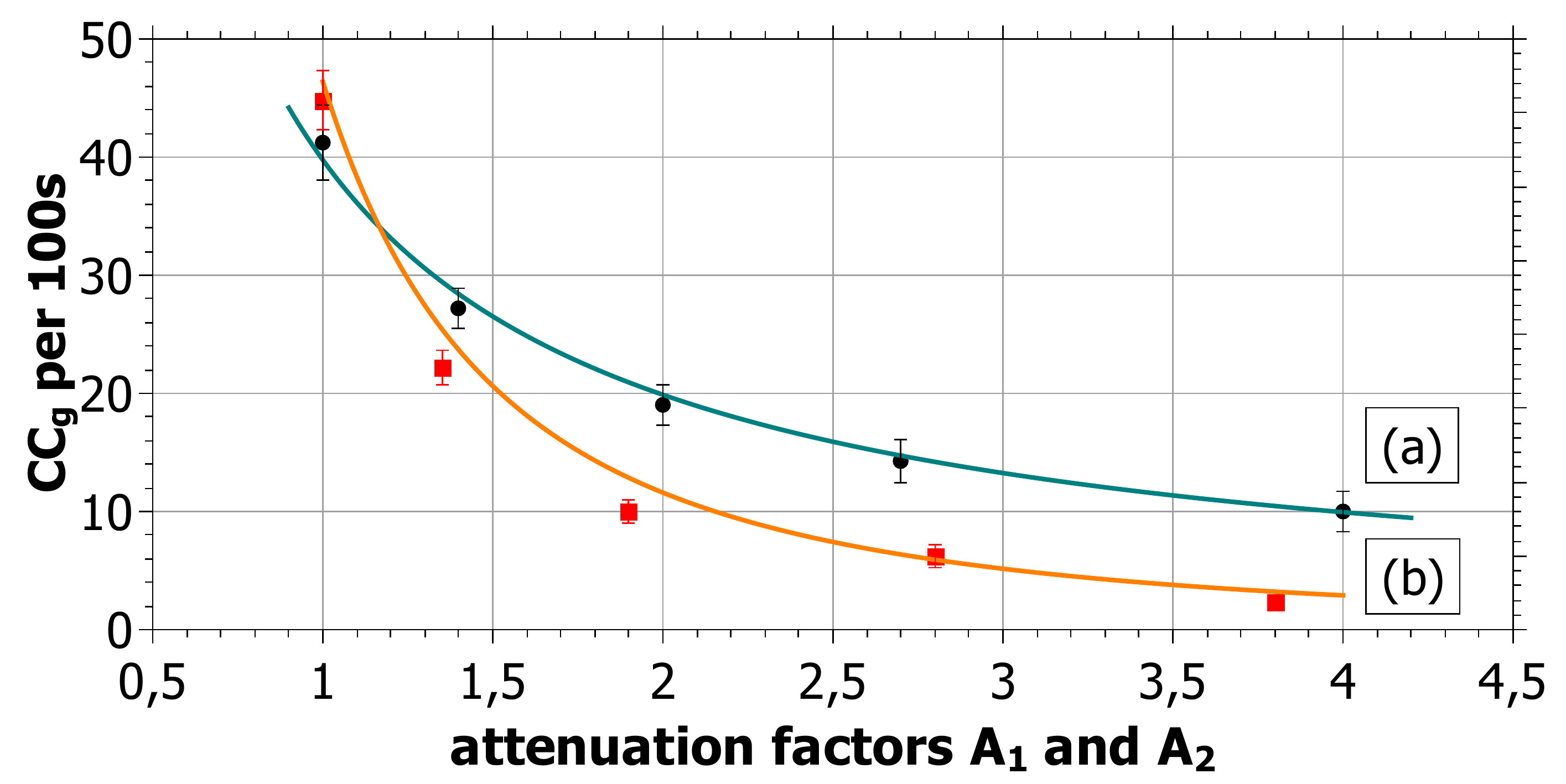}
		\caption{(a) Dependence of $CC_{g}$ on attenuation factor $A_{1}$. Points visualize experimentally observed results, the solid blue--green line depicts fitted experimental data with theoretical fit based on Eq. (\ref{eq:SNR_CCg}). (b) Dependence of $CC_{g}$ on attenuation factor $A_{2}$. The solid orange line depicts fitted experimental data with theoretical dependence based on Eq. (\ref{eq:SNR_CCg}).}	
		\label{attenuation}	
		\end{center}
\end{figure}

The final two tests of our model involved verifying the dependence of genuine coincidence rate $CC_g$ on the coupling efficiencies (i) $t_1$ and (ii) $t_2$ as predicted in Eq. (\ref{eq:SNR_CCg}). During each of the two tests, the parameter $R$ and the pumping power were kept constant resulting in constant SNR. During the first test the value of SNR was \SI[separate-uncertainty]{4.7(16)}{\decibel}. In the second test the SNR was \SI[separate-uncertainty]{5.0(13)}{\decibel}. In order to test the dependence on idler and signal mode transmissivities $t_1$ and $t_2$, we have acquired the coincidences in the usual three steps for various levels of attenuation by closing a diaphragm on the idler and signal mode fiber couplers respectively. When the signal mode attenuation was set, the NDF3 in the attenuated fundamental laser mode was readjusted to maintain a constant $R$. This was not necessary when closing the idler mode diaphragm. For better readability of our results, we introduce the idler and signal mode intensity attenuation factors $A_1$ and $A_2$ so that the modes' transmissivities become $t_{j}^2 \rightarrow t_{j}^2/A_{j}$ for $j=1,2$. Experimentally observed values are summarized in Tab. \ref{tabatten} and visualized in Fig. \ref{attenuation}. Fig. \ref{attenuation} demonstrates that with constant SNR $CC_{g}$ dependents on modes' transmissivities $t_{1}^2$ and $t_{2}^2$ as functions $\frac{1}{x}$ and $\frac{1}{x^2}$ respectively as predicted in Eq.(\ref{eq:SNR_CCg}).

\begin{table}
\begin{ruledtabular}
\begin{tabular}{cc|cc}
\multicolumn{2}{ c }{\bfseries idler attenuation (t${}_1$)}& \multicolumn{2}{ c }{\bfseries signal attenuation (t${}_2$)}\\
\hline
    \bfseries A${}_1$ & \bfseries CC${}_g$ per \SI{100}{\second} & \bfseries A${}_2$ & \bfseries CC${}_g$ per \SI{100}{\second} \\[2mm]
    \hline
    1 &41.2 $\pm$ 3.2 & 1 & 44.8  $\pm$ 2.5 \\
    1.4 & 27.2 $\pm$ 1.7 & 1.3 & 22.2  $\pm$ 1.5 \\
    2 & 19.0 $\pm$ 1.7 & 1.9 & 10.0  $\pm$ 1 \\
    2.7 & 14.3 $\pm$ 1.8 & 2.8 & 6.2  $\pm$ 1 \\
    4 & 10.0 $\pm$ 1.7 & 3.8 & 2.3  $\pm$ 0.3 \\
\end{tabular}
\caption{Experimentally observed data and their respective errors when investigating the dependence of $CC_{g}$ on the attenuation factors $A_{1}$ and $A_{2}$.}
\label{tabatten}
\end{ruledtabular}
\end{table}

\section{Impact of the noise on teleportation fidelity}
\label{sec4}
%\emph{Impact of the noise on teleportation fidelity} --- 
We now investigate the impact of the above analyzed noise on quantum teleportation. Since quantum teleportation is a key ingredient in many quantum information protocols, it is essential to asses the influence of inherent noise of various photon sources on its performance. In quantum circuits, including teleporation, one often uses fidelity as a measure of the circuits quality. Assuming a pure input qubit state $|\psi\rangle_{in}$ and the resulting teleported state $\hat{\rho}_{out}$, fidelity can be calculated using the formula

\begin{equation}
F = |\langle \psi_{in}|\hat{\rho}|\psi_{in}\rangle|.
\end{equation}

Note that when teleportation is replaced by classical ``measure and recreate'' protocol, the fidelity can not exceed its classical limit of $\frac{2}{3}$ \cite{fidelity}. Even though it is impossible to reach perfect fidelity $F=1$ in realistic conditions, one still targets to maximize its value.

In our analysis we have calculated the dependence of average fidelity $\langle F \rangle$ on the signal-to-noise ratio (SNR). If we fix the parameter $R$ to its optimum value ($R \approx 0.35$) the fidelity $\langle F \rangle$ is than a function that depends on $CC_{g}$ and only one of the $CC_{s}$ or $CC_{f}$ since these two are bound by fixed parameter $R$. As a result the fidelity is a function of SNR. We have calculated the average fidelity using the formula
\begin{equation}
\langle F \rangle = \dfrac{P_{CC_{g}} F_{g} + P_{CC_{s}} F_{s} + P_{CC_{f}} F_{f}}{P_{CC_{g}} + P_{CC_{s}} + P_{CC_{s}}},
\label{eq:fidelity}
\end{equation}
where
\begin{equation}
P_{CC_{g}} = \dfrac{CC_{g}}{4f}, P_{CC_{s}} = \dfrac{CC_{s}}{4f}, P_{CC_{f}} = \dfrac{CC_{f}}{4f},
\end{equation}
are the probabilities of the coincidence events. $f$ stands for the repetition rate of the pumping laser and $F_{g}$, $F_{s}$, $F_{f}$ are the teleportation fidelities if the coincidence $CC_{g}$, $CC_{s}$ or $CC_{f}$ occur respectively. The value of teleportation fidelity $F_{g} = 1$ because from the definition there is one photon in each mode so the teleportation succeeds perfectly, at least in principle. On the other hand, the teleportation fidelities $F_{s}$ and $F_{f}$ have values of $\frac{1}{2}$. First one because the two photons in signal mode are randomly projected onto Bell states uncorrelated with the teleported photon which is missing. The later because the two photons in attenuated laser mode are not correlated with the idler mode which is thus a mixed state.

\begin{figure}[!htb]
		\begin{center}
		\includegraphics[scale=0.3]{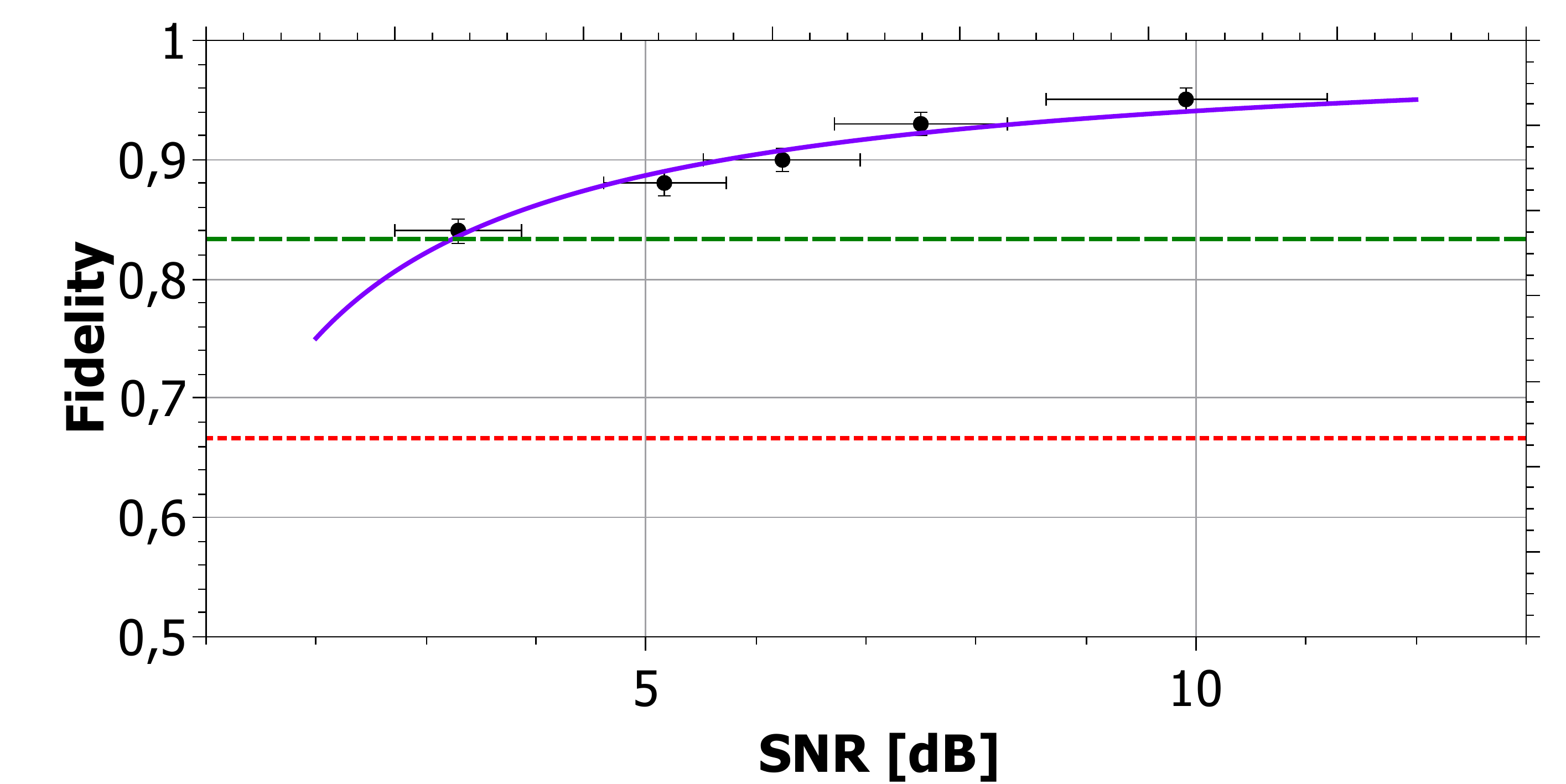}
		\caption{Dependence of average fidelity $\langle F \rangle$ on SNR. Points visualize calculated results from experimentally observed SNRs. {The solid violet line  corresponds to our theoretical model, the dotted red line is the classical protocol limit ($F = 2/3$)\cite{fidelity} and the dashed green line indicates the secure teleportation, i.e., $F=5/6$ cloning threshold see \cite{Sahin}.}}	
		\label{fidelity}	
		\end{center}
\end{figure}

Calculated values are summarized in Tab. \ref{tab3} and visualized in Fig. \ref{fidelity}. We observe that the average fidelity drops only slightly with decreasing SNR, so the average fidelity is above 80\% for SNR around \SI{3}{\decibel}. However this does not take into account other experimental imperfections (such as two--photon overlap, polarization adjustments etc.) that combining with photon-number noise can lead to such a low fidelity that the protocol fails. The fidelity uncertainty intervals were calculated using a Monte--Carlo simulation based on poisson distribution of detected coincidences.

\begin{table}
\begin{ruledtabular}
\begin{tabular}{ccc}
    \bfseries  fidelity F & \bfseries fidelity uncertainty interval & \bfseries SNR [dB]\\[2mm]
    \hline
    0.96 & $\langle 0.93,0.98 \rangle$ &9.91 $\pm$ 1.27  \\
    0.94 & $\langle 0.90,0.96 \rangle$ &7.50 $\pm$ 0.79  \\
    0.92 & $\langle 0.86,0.95 \rangle$ &6.23 $\pm$ 0.71  \\
    0.89 & $\langle 0.85,0.91 \rangle$ &5.17 $\pm$ 0.56  \\
    0.85 & $\langle 0.83,0.86 \rangle$ &3.29 $\pm$ 0.58  \\
\end{tabular}
\caption{Calculated data and their respective errors when investigating the dependence of average fidelity F on the $SNR$.}
\label{tab3}
\end{ruledtabular}
\end{table}

\section{Conclusions}
\label{conclusions}
In conclusion, we have shown that our model fits the experimental data very well. We have demonstrated the role of the ratio $R$ between the SPDC-based and attenuated fundamental-based false coincidences. We have also confirmed its optimal value being close to 1 depending on the pumping strength. In the next step, we have verified that SNR (when optimal $R$) can only be increased by decreasing the SPDC pumping strength. Our data fit well both the SNR as a function of genuine coincidence rate, and also the predicted coincidence rate as a function of pumping strength. Finally, we have successfully tested the genuine coincidence rates as functions of coupling efficiencies while maintaining constant SNR. Our model and the obtained conclusions drawn from it can be useful for experimentalist when constructing a similar three-photon source and using it for teleportation-like protocols. With respect to that, we have made a prediction of the impact of this noise to teleportation fidelity. While fidelity drops smoothly with decreasing SNR,  in conjunction with other experimental imperfections may lead to fidelity below the classical threshold.

\section*{Acknowledgement}
Authors acknowledge
financial support by the Czech Science Foundation under the project No. 17-10003S. KL and KB also acknowledge the financial support 
of the Polish National Science Centre under grant
DEC-2013/11/D/ST2/02638.
The authors also acknowledge the project
No. LO1305 of the Ministry of Education, Youth and
Sports of the Czech Republic financing the infrastructure of their workplace and VT also acknowledges the Palacky University internal grant No. IGA-PrF-2017-005

\end{document}